%% file: main.tex
\documentclass[sigconf]{acmart}
\usepackage{graphicx} 
\usepackage{float}

\copyrightyear{2026}
\acmYear{2026}
\setcopyright{cc}
\setcctype{by}
\acmConference[ISS Companion '26]{Companion Proceedings of the Conference on Interactive Surfaces and Spaces}{November 23--26, 2026}{Turin, Italy}
\acmBooktitle{Companion Proceedings of the Conference on Interactive Surfaces and Spaces (ISS Companion '26), November 23--26, 2026, Turin, Italy}
\acmDOI{10.1145/3845796.3859295}
\acmISBN{979-8-4007-3011-5/2026/11}
\begin{document}

\title{TableControl: Gesture Interaction Through Surface Vibrations}
\author{Florian Hettstedt}
\affiliation{
    \institution{University of Duisburg-Essen}
    \city{Duisburg}
    \country{Germany}
}
\author{Andreas Erbslöh}
\affiliation{
    \institution{University of Duisburg-Essen}
    \city{Duisburg}
    \country{Germany}
}
\author{Gregor Schiele}
\affiliation{
    \institution{University of Duisburg-Essen}
    \city{Duisburg}
    \country{Germany}
}

\input{body/00_abstract}
\maketitle
\raggedbottom

\input{body/01_introduction}
\input{body/02_related_work}
\input{body/03_system_overview}
\input{body/04_interactive_demonstration}
\input{body/05_discussion_and_conclusion}

\bibliographystyle{ACM-Reference-Format}
\bibliography{reference}
\end{document}

%% file: body/00_abstract.tex
\begin{abstract}
Integrating gesture sensing into furniture offers a way to interact with digital applications and physical devices through familiar surfaces. We present TableControl, an interactive demonstration of continuous surface-vibration-based gesture recognition with the complete processing pipeline running on a microcontroller. Using four concealed piezoelectric sensors, the system combines signal acquisition and preprocessing with neural event detection and gesture classification. The resulting labels for knock, swipe-up, and swipe-down gestures are transmitted to a host computer without streaming the underlying sensor signals. Visitors explore two example applications through hands-on interaction: an endless runner game and a physical table lamp with on/off and brightness control. A dashboard displays recognized gestures, class probabilities, and processing times. 

\end{abstract}
\ccsdesc[300]{Human-centered computing~Gestural input}
\keywords{surface vibration, gesture recognition, smart furniture, tabletop interaction, sensor data processing, embedded deep learning}

%% file: body/01_introduction.tex
\section{Introduction}
\label{sec:introduction}
Smart-home technologies offer new ways for people to interact with their living environments. This type of interaction is commonly based on smartphone applications, dedicated control panels, or voice assistants. A complementary and unobtrusive approach is smart furniture, where input capabilities are integrated into existing objects. Therefore, familiar surfaces can become interfaces while retaining their appearance and primary function.

Surface-vibration sensing provides one way to realize such interfaces. Knocks and swipes generate mechanical vibrations that can be captured by sensors mounted underneath the surface and analyzed for gesture recognition. \textit{Smatable}~\cite{smatable2023} demonstrated this approach by recognizing swipe gestures on ordinary tables using concealed vibration sensors. Building on this approach, Hettstedt et al.~\cite{hettstedt2026} introduced a custom sensing platform, a gesture dataset, and a configurable preprocessing and classification pipeline, demonstrating the feasibility of recognizing surface gestures from recorded vibration signals.

Extending this work, we present TableControl, an interactive demonstration of surface-vibration-based gesture recognition in which the complete sensor data processing pipeline is executed on a microcontroller. Signal acquisition and preprocessing, including filtering and windowing, are combined with a two-stage neural recognition pipeline consisting of an event detector and a gesture classifier. The system detects and classifies gestures and transmits the resulting labels to a host computer for application control without streaming the underlying sensor signals.

To demonstrate the system's functionality through hands-on interaction, we present two example applications: an endless runner game and the control of a physical table lamp. The game provides a playful setting in which users can experience how the system responds to their gestures, while lamp control illustrates everyday smart-home interaction. A dashboard displays recognition outputs and processing times. Together, these examples illustrate how gestures performed on a tabletop can be used to control both digital applications and physical devices.
\begin{figure*}[t]
    \centering
    \includegraphics[width=0.98\linewidth]{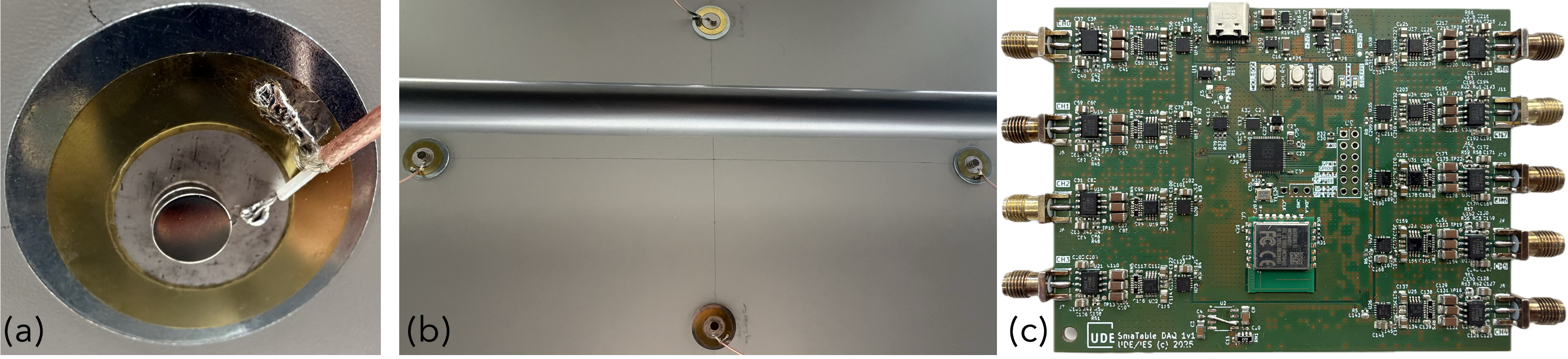}
    \caption{Sensing hardware: (a) close-up of a sensor unit,
    (b) four sensors mounted beneath the tabletop, and (c) acquisition board.}
    \Description{Three photographs showing the table setup,
    the acquisition board, and a close-up of a sensor unit.}
    \label{fig:combi-fig-hardware}
\end{figure*}


%% file: body/02_related_work.tex
\section{Related Work}
\label{sec:related_work}
Previous research has explored ways to turn everyday objects and surfaces into input interfaces, each with specific requirements for integrating the interface into its environment. \textit{HIC}~\cite{HIC} provides a projected graphical interface for controlling smart-home appliances, requiring projection hardware in the interaction space. \textit{Sensurfaces}~\cite{sensurfaces} integrates modular capacitive sensing into surfaces made from materials such as wood and concrete while preserving their appearance. However, it requires an underlying electrode array, and its sensitivity depends on the material's thickness and dielectric properties. \textit{Acustico}~\cite{acustico} leaves the surface uninstrumented by using wrist-worn microphones and accelerometers to detect and localize taps, but requires users to wear a sensing device. For TableControl, we pursue gesture input through an existing tabletop using concealed sensors attached to the furniture by placing the sensor elements and the hardware beneath the tabletop.

Surface-vibration sensing provides a basis for this approach by capturing mechanical signals generated during interaction. \textit{SAWSense}~\cite{SAWSense} demonstrates how voice pickup units attached to objects can recognize touch gestures and activities from surface acoustic waves. \textit{VibSense}~\cite{VibSense} uses piezoelectric sensors to localize touches and, with active vibration excitation, identify objects. Although TableControl also uses piezoelectric sensing, its interaction model relies on distinguishing knock and swipe gestures rather than identifying contact locations or objects. More directly aligned with this goal, \textit{Smatable}~\cite{smatable2023} combines concealed piezoelectric sensors beneath ordinary tables with a spectrogram-based two-dimensional convolutional neural network~(CNN) to classify swipe directions.~

Building on this sensing approach, Hettstedt et al.~\cite{hettstedt2026} introduced the sensing platform and gesture dataset used in this work and investigated preprocessing and compact gesture-classification models. Their evaluation focused on recorded gesture data, with deployment of the complete processing pipeline identified as future work. TableControl addresses this next step by integrating signal acquisition, preprocessing, neural event detection, and gesture classification into a continuous pipeline running entirely on a microcontroller. The demonstration uses the resulting gesture events to control both an interactive game and a physical table lamp.
\begin{figure*}[t]
    \centering
    \includegraphics[width=0.95\linewidth,
    pagebox=cropbox,
    clip]
        {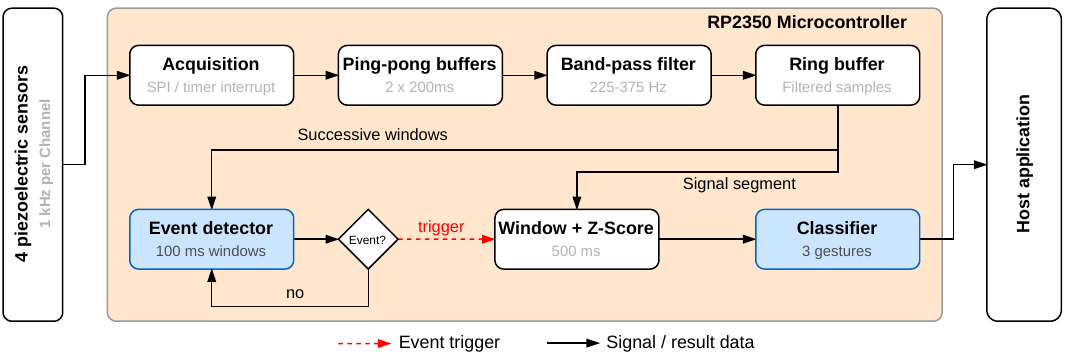}
    \caption{Continuous on-device gesture recognition pipeline.
    A shared ring buffer provides the input windows for event
    detection and gesture classification. The dashed arrow
    indicates the event trigger. Recognition results are
    transmitted to the host application.}
    \Description{Four piezoelectric sensors provide signals that
    pass through acquisition, alternating buffers, and a band-pass
    filter into a ring buffer. The event detector evaluates
    100 ms windows. An accepted detection triggers classification
    once the corresponding 500 ms window is available. This window
    is normalized before classification. Recognition results
    are sent to the host application.}
    \label{fig:recognition-pipeline}
\end{figure*}

%% file: body/03_system_overview.tex
\section{System Overview}
\label{sec:system_overview}
TableControl consists of a concealed sensing setup, a continuous on-device recognition pipeline, and host software for game and lamp control.

\subsection{Sensing Hardware}
The microcontroller enables the local inference of the whole recognition pipeline, transmitting recognition results to the host computer for application control.
TableControl uses the custom sensing platform introduced by Hettstedt et al.~\cite{hettstedt2026}. Figure~\ref{fig:combi-fig-hardware} shows (a) a close-up of a sensor unit, (b) the concealed sensor placement on the underside of the tabletop, and (c) the custom acquisition hardware combining the analog front-end to record the continuous sensor data stream and the computing unit.
Four sensor units are mounted underneath the tabletop to capture mechanical vibrations generated by gestures on its surface.
Their concealed placement preserves the appearance of the tabletop and leaves the interaction area unobstructed. 
Each sensor unit combines a piezoelectric element with two small neodymium magnets and is mounted beneath the tabletop using a self-adhesive metal disk.
Shielded cables connect the sensors to the analog front-end of the custom acquisition board. The board provides analog signal processing and 18-bit analog-to-digital conversion for up to nine channels and incorporates an RP2350 microcontroller. The signals from the four sensors are acquired at 1~kHz per channel and passed to the embedded processing pipeline running on the microcontroller.

\subsection{On-Device Gesture Recognition}
Figure~\ref{fig:recognition-pipeline} provides an overview of the recognition pipeline, which runs continuously on the microcontroller. It follows a two-stage design: an event detector identifies potential gestures, and a gesture classifier determines their class. Shared signal acquisition and preprocessing provide the input data for both stages.

The microcontroller acquires digitized sensor signals via SPI using a timer interrupt and stores the samples in two alternating buffers, each holding 200~ms of data. When one buffer is full, acquisition switches to the other while the completed block is processed. 
The duration of the block was selected to accommodate the measured classifier inference time while leaving additional time for preprocessing, event detection, and result transmission. To avoid buffer overruns, processing of the completed block must finish before the other buffer fills.
Samples from each completed block are filtered independently for each channel using a second-order Butterworth IIR-bandpass filter within the range of [225~Hz, 375~Hz]. Filter states are preserved across buffer boundaries to maintain continuous filtering. The filtered samples are written to a separate ring buffer that retains signal history across successive blocks. This history remains available after the corresponding acquisition buffer has been processed and released, providing the input windows for both recognition stages.

In the first stage, the event detector evaluates successive 100~ms windows of filtered samples from the ring buffer with a stride of 100~ms and distinguishes between gesture events and background activity. These windows are evaluated sequentially as the completed acquisition block is processed. An accepted detection triggers the second stage, with the midpoint of the detector window serving as the estimated event timestamp.
The second stage uses a 500~ms classification window, covering 50~ms before and 450~ms after the estimated event time. Acquisition and filtering continue until the complete window is available in the ring buffer. The window is then extracted, normalized using channel-wise Z-score normalization, and passed to the gesture classifier, which distinguishes between three gestures: knock, swipe-up, and swipe-down. While an event is awaiting classification, additional detections are not accepted. The larger classifier therefore runs only for segments associated with an accepted event.
Both deep neural networks use depthwise separable one-dimensional convolutions, combining depthwise temporal filtering with pointwise convolutions across feature channels. The event detector comprises three convolutional blocks with a kernel size of 13, while the gesture classifier uses six blocks with a kernel size of 21. Each block includes batch normalization, \textit{ReLU} activation, and \textit{max pooling} with a factor of 2. Global average pooling is followed by a fully connected layer with 16 units, \textit{ReLU} activation, and an output layer with \textit{softmax}. The resulting class probabilities can be used to reject predictions whose maximum class probability falls below a chosen threshold. The detector contains 2,198 and the classifier 8,311 trainable parameters.

The event detector and gesture classifier were trained on the knock, swipe-up, and swipe-down subset of the dataset introduced by Hettstedt et al.~\cite{hettstedt2026_dataset}. While the original recordings were collected in a single laboratory environment, everyday use can involve additional background vibrations induced by environmental sounds. These vibrations are captured by the sensors and can interfere with both event detection and gesture classification. To expose the models to disturbances beyond the original recording conditions, training included clean recordings and versions additively augmented with separately recorded sensor signals from music, conversations, and podcasts at nominal signal-to-noise ratios (SNRs) of [20, 10, 5, 0, $-5$, $-10$]~dB. 

Noise scaling was based on mean power over unfiltered 1~s event windows and complete noise recordings. For event detection, 100~ms band-pass filtered windows were extracted around annotated gesture onsets, with uniformly sampled temporal shifts within $\pm50$~ms during training. Each condition included an equal number of negative windows from separate quiet or noise-only recordings, using the same noise scaling as the corresponding positive recordings for augmented conditions. Each of the 19 training conditions contained 4,500 gesture windows. Including the additional background windows for event detection, this yielded 85,500 training samples per epoch for the gesture classifier and 171,000 for the event detector.

Both networks execute locally using \textit{TensorFlow Lite Micro} with 32-bit floating-point operations. The preprocessing methods are generated with the \textit{elasticAI.preprocessor} framework\footnote{Access to \textit{elasticAI.preprocessor} framework via GitHub: \url{https://github.com/es-ude/elasticai.preprocessor}}.  Signal acquisition, preprocessing, event detection, and gesture classification therefore remain on the microcontroller. The resulting gesture label and class probabilities are transmitted to the host computer for application control.

In an offline evaluation using leave-one-subject-out cross-validation with clean and noise-augmented samples, the event detector and gesture classifier achieved mean accuracies of 82.5\% and 85.7\%, respectively. In each fold, all samples from one participant were reserved for testing, while the models were trained on samples from the remaining participants.
Across ten inference calls per neural network, mean inference times were 14.7~ms for the event detector and 107.1~ms for the gesture classifier. The mean latency from the estimated event timestamp to the completion of gesture classification was 598.2~ms. This interval includes buffering and classification-window collection, but excludes transmission to the laptop and application rendering.

\subsection{Application Interface}
The acquisition board transmits recognition results to the host computer via USB serial or TCP over Wi-Fi, depending on the firmware configuration. Each recognized gesture produces a newline-terminated JSON message containing the gesture identifier, class probabilities, the estimated event timestamp, the inference times of both neural networks, and the event-to-classification latency.
A Python bridge running on the host receives and validates these messages and forwards the recognition results to the application via local UDP messages. The bridge provides a common interface for both transport variants, allowing the application to consume gesture events independently of the connection to the acquisition board.

\begin{figure*}[t]
    \centering
    \includegraphics[width=0.98\linewidth]
        {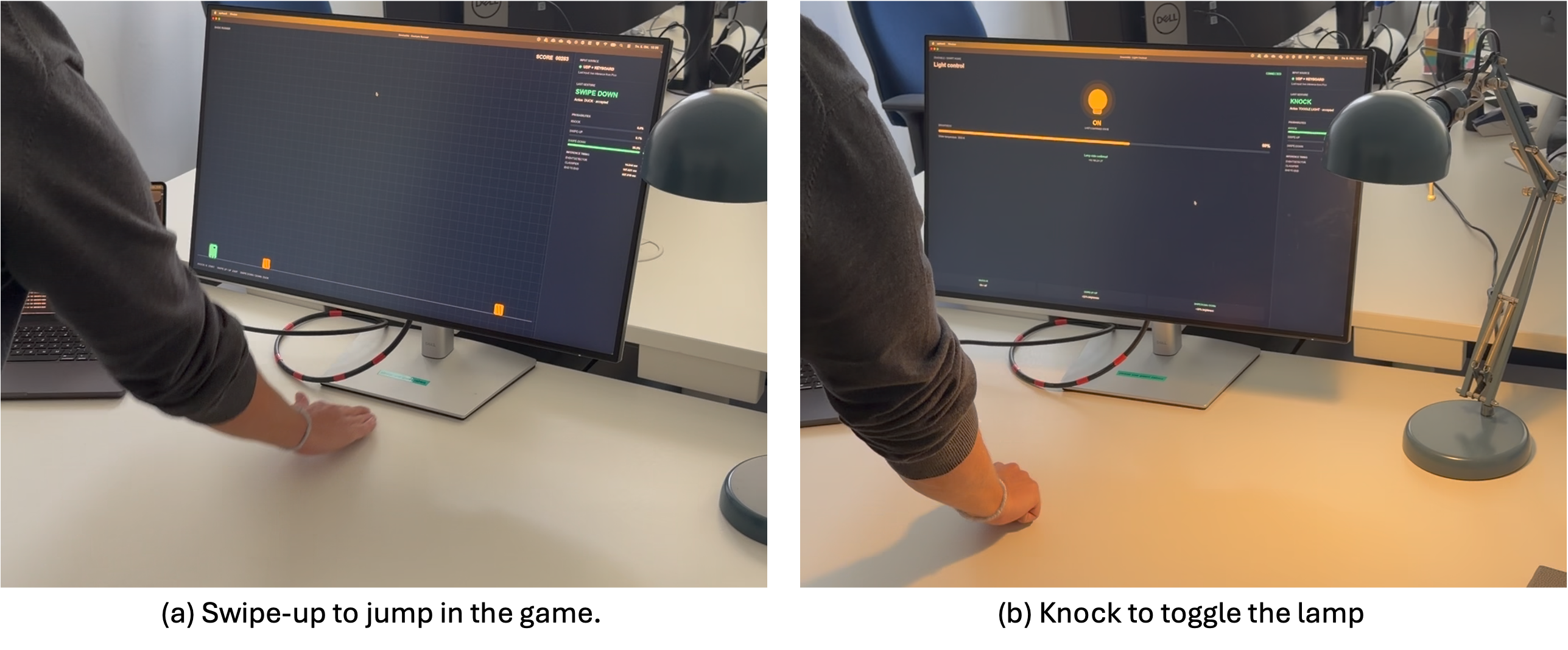}
    \caption{TableControl interaction examples: 
    (a) Swipe-up to jump in the game;
    (b) Knock to toggle the lamp.}
    \Description{Two side-by-side photographs show gesture interaction on an instrumented table. Left: a visitor's open hand touches the tabletop i front of a monitor displaying the endless runner game and recognition dashboard. Right: a visitor's fist touches the tabletop beside an illuminated lamp. The monitor displays the lamp-control interface and recognition dashboard.}
    \label{fig:demo-interface}
\end{figure*}

%% file: body/04_interactive_demonstration.tex
\section{Interactive Demonstration}
\label{sec:interactive_demonstration}
TableControl allows visitors to control either an endless runner game or a physical table lamp by performing gestures on an instrumented tabletop. The demonstration setup includes a laptop, the lamp and everyday objects placed on the table. The central area remains clear for gesture interaction, as the sensors are mounted directly beneath it.

In the endless runner game, implemented in Python using \textit{pygame-ce}, the character runs automatically. Visitors use swipe-up gestures to jump over low obstacles (Figure~\ref{fig:demo-interface}a) and swipe-down gestures to duck beneath elevated obstacles. A knock starts a short tutorial that introduces the controls through two jumps and two ducking actions without obstacles. Once these actions have been completed, the game begins automatically. A collision ends the run, and another knock resets the game and score for a new attempt.

The second application uses the same gestures to control a physical table lamp. A knock switches the lamp on or off (Figure~\ref{fig:demo-interface}b), while swipe-up gestures increase its brightness and swipe-down gestures decrease it. The host computer sends the corresponding control commands to the lamp over Wi-Fi, with the resulting changes directly visible at the lamp. This application demonstrates how gestures performed on the tabletop can serve as input for an everyday physical device.

The game and lamp provide visible feedback on the triggered actions, while a recognition dashboard provides additional information about how the system interprets the gestures. It displays the most recent recognized gesture and the probabilities assigned to the gesture classes, alongside the inference times of both neural networks and the event-to-classification latency.



%% file: body/05_discussion_and_conclusion.tex
\section{Discussion and Conclusion}
\label{sec:discussion_and_conclusion}
TableControl demonstrates how tabletop gestures can control digital applications and physical devices, with recognition performed entirely on a microcontroller. The game and lamp scenarios illustrate how the same gestures can support playful interaction and everyday device control, highlighting the potential of familiar furniture surfaces as input interfaces.

Future work will investigate shorter classification windows and model quantization to reduce latency and resource usage while assessing their effects on recognition performance. Further evaluation will examine recognition across users, tabletop materials, object arrangements, and environmental vibrations, with particular attention to unintended activations during everyday table use. We will also investigate gesture mappings and application feedback to guide the design of further applications for smart furniture.



%% file: reference.bib
@ARTICLE{smatable2023,
  author={Yoshida, Makoto and Matsui, Tomokazu and Ishiyama, Tokimune and Fujimoto, Manato and Suwa, Hirohiko and Yasumoto, Keiichi},
  journal={IEEE Access}, 
  title={Smatable: A Vibration-Based Sensing Method for Making Ordinary Tables Touch-Interfaces}, 
  year={2023},
  volume={11},
  number={},
  pages={142611-142627},
  doi={10.1109/ACCESS.2023.3343500}
  }

@INPROCEEDINGS{hettstedt2026,
  author={Hettstedt, Florian and Giese, Cedric and Ling, Tianheng and Yasumoto, Keiichi and Schiele, Gregor and Erbslöh, Andreas},
  booktitle={IEEE Int. Conf. on Pervasive Computing and Communications Workshops (PerCom Workshops)}, 
  title={Towards an End-To-End System for Real-Time Gesture Recognition from Surface Vibrations}, 
  year={2026},
  volume={},
  number={},
  publisher={IEEE},
  address={Piscataway, NJ, USA},
  pages={1-6},
  doi={10.1109/PerComWorkshops68308.2026.11585250}
}

@dataset{hettstedt2026_dataset,
  author       = {Hettstedt, Florian and
                  Giese, Cedric and
                  Ling, Tianheng and
                  Yasumoto, Keiichi and
                  Schiele, Gregor and
                  Erbslöh, Andreas},
  title        = {Data for: Towards an End-To-End System for Real-
                   Time Gesture Recognition from Surface Vibrations
                  },
  month        = may,
  year         = 2026,
  publisher    = {Zenodo},
  doi          = {10.5281/zenodo.20408458},
  url          = {https://doi.org/10.5281/zenodo.20408458},
}

@inproceedings{acustico,
    author = {Gong et al., Jun},    
    title = {{Acustico: Surface Tap Detection and Localization using Wrist-based Acoustic TDOA Sensing}},
    year = {2020},
    doi = {10.1145/3379337.3415901},
    booktitle = {33rd Annual ACM Symp. on User Interface Software and Technol.},
    pages = {406–419},
    publisher={ACM},
    address={New York, NY, USA},
    numpages = {14},
}

@inproceedings{SAWSense,
    author = {Iravantchi et al., Yasha},
    title = {{SAWSense: Using Surface Acoustic Waves for Surface-bound Event Recognition}},
    year = {2023},
    doi = {10.1145/3544548.3580991},
    booktitle = {Conf. on Human Factors in Computing Systems},
    articleno = {422},
    publisher={ACM},
    address={New York, NY, USA},
    numpages = {18},
}

@article{sensurfaces,
    author = {Parilusyan et al., Brice},
    title = {{Sensurfaces: A Novel Approach for Embedded Touch Sensing on Everyday Surfaces}},
    year = {2022},
    publisher = {Association for Computing Machinery},
    address = {New York, NY, USA},
    volume = {6},
    number = {2},
    doi = {10.1145/3534616},
    journal = {Proc. ACM Interact. Mob. Wearable Ubiquitous Technol.},
    month = jul,
    articleno = {67},
    numpages = {19}
}

@inproceedings{VibSense,
    author = {Liu et al., Jian},
    title = {{VibSense: Sensing Touches on Ubiquitous Surfaces through Vibration}},
    year = {2017},
    publisher = {IEEE},
    address={Piscataway, NJ, USA},
    doi = {10.1109/SAHCN.2017.7964907},
    booktitle = {14th Annual IEEE Int. Conf on Sensing, Communication, and Networking},
    pages = {1–9},
}

@INPROCEEDINGS{HIC,
    author={Pizzagalli et al., Simone},
    booktitle={IEEE 6th Int. Conf. on Serious Games and Applications for Health}, 
    title={{HIC: An interactive and ubiquitous home controller system for the smart home}}, 
    year={2018},
    volume={},
    number={},
    pages={1-6},
    publisher={IEEE},
    address={Piscataway, NJ, USA},
    doi={10.1109/SeGAH.2018.8401374}
}
